\documentclass[preprints,communication,accept,pdftex,moreauthors]{Definitions/mdpi} 

\firstpage{1} 
\pubvolume{1}
\issuenum{1}
\articlenumber{0}
\pubyear{2023}
\copyrightyear{2023}
\datereceived{ } 
\daterevised{ } 
\dateaccepted{ } 
\datepublished{ } 
\hreflink{https://doi.org/} 
\pdfoutput=1 

\usepackage{babel}[english]

\Title{Event-shape-dependent analysis of charm-anticharm azimuthal correlations in simulations}

\TitleCitation{Event-shape-dependent analysis of charm-anticharm correlations in simulations}

\Author{Anikó Horváth $^{1,2}\orcidA{}$, Eszter Frajna$^{1,3}\orcidB{}$ and Róbert Vértesi$^{1}\orcidC{}$}

\AuthorNames{Anikó Horváth, Eszter Frajna and Róbert Vértesi}

\AuthorCitation{Horváth, A.; Frajna, E.; Vértesi, R.}

\address{%
$^{1}$ \quad Wigner Research Centre for Physics, P.O. Box 49, H-1525 Budapest, Hungary\\
$^{2}$ \quad Eötvös Loránd University, Budapest, Hungary\\
$^{3}$ \quad Budapest University of Technology and Economics, Hungary}

\corres{Correspondence: aniko.horvath@wigner.hu, vertesi.robert@wigner.hu}

\abstract{
In high-energy collisions of small systems, by high-enough final-state multiplicities, a collective behaviour is present that is similar to the flow patterns observed in heavy-ion collisions. Recent studies connect this collectivity to semi-soft vacuum-QCD processes. Here we explore QCD production mechanisms using angular correlations of heavy flavour using simulated proton-proton collisions at $\sqrt{s} = 13$~TeV with the PYTHIA8 Monte Carlo event generator. We demonstrate that the event shape is strongly connected to the production mechanisms. Flattenicity, a novel event descriptor, can be used to separate events containing the final-state radiation from the rest of the events.
}


\begin{document}

\section{Introduction}

In high-energy heavy-ion collisions, a strongly interacting quark-gluon plasma (QGP) is created, which was found to behave as an almost-perfect fluid~\cite{PHENIX:2004vcz,STAR:2005gfr}. Surprisingly, a similar collective behavior was observed in small (proton--proton and proton--nucleus) collisions with high final-state multiplicity~\cite{CMS:2010ifv,ALICE:2019zfl}. Whether or not QGP is created in these smaller collision systems is still an open question today. Recent works suggest that vacuum-QCD processes on the soft-hard boundary, such as multiple-parton interactions (MPI) with colour reconnection, are able to generate the collective patterns that are observed in such systems~\cite{Bierlich:2017vhg,Ortiz:2016kpz}.

Heavy quarks are mostly created in the early stages of the collision, in perturbatively accessible quantum-chromodynamics (QCD) processes: a heavy quark can be created from a pair of gluons or light quarks by flavour creation (FLC), a gluon splitting into the quark-antiquark pair (GSP), or through flavour excitation (FLX)~\cite{Norrbin:2000zc,Ilten:2017rbd}. Moreover, they may interact in semi-hard processes and participate in the formation of the underlying event~\cite{Vertesi:2021ihp}.
Further insight to the connection of the hard process and the underlying event can be gained by the differential exploration of events with respect to event-shape variables. While traditionally the final-state multiplicity is used to categorize events by activity, other recently introduced event-shape variables, such as transverse spherocity and flattenicity, are sensitive to event topology and have a more direct connection to multiple-parton interactions and the emerging collective patterns~\cite{Ortiz:2022zqr,Ortiz:2015ttf}.

Angular correlation measurements are sensitive probes of parton production and fragmentation down to low momenta where jet reconstruction is problematic in a rich final-state environment. The current experimental precision allows for the exploration of heavy-flavour hadron correlations, which provides information about heavy-flavour fragmentation but very little insight to their creation. A recent ALICE measurement did not find event-activity dependence in the angular correlation of D$^0$ mesons to charged hadrons~\cite{ALICE:2021kpy}. The experimental possibilities will be significantly extended with the arrival of LHC Run3 data, where heavy-flavour--heavy-flavour correlations will be possible to reconstruct. The full potential of the forthcoming LHC Run3+Run4 data, however, can be exploited with measurements that are differential in event-activity or event-shape.

In this work we explore the azimuthal correlations of charm--anticharm quark pairs in Monte-Carlo (MC) simulations, in terms of different event-activity variables. 
We use MC information to explore the connection of the partonic processes with the emerging final state and propose an experimental method to separate them.

\section{Methods}
We analyzed proton--proton collisions simulated at $\sqrt{s} = 13$ TeV center-of-mass energy with the PYTHIA8~\cite{Sjostrand:2007gs} (version 8.308) MC event generator. 
%
In PYTHIA, the initial leading-order production process is amended by other partonic processes: initial and final-state radiation (ISR and FSR respectively) as well as multiple-parton interactions.
By enabling these processes one by one, we simulated just the initial hard process (all off), the initial hard scattering and multiparton interactions added (MPI on), with the initial-state gluon radiations included (MPI, ISR on), and with all the previous processes and the final-state radiations also enabled in the events (all on).
With each setting, 10 million events were simulated.

We computed the azimuthal correlations of charm quarks with anticharm quarks. Only those quarks that directly hadronised were considered, to avoid multiple counting of the same quark. As an arbitrary choice, charm quarks were used as trigger particles and anticharm quarks as associated particles. Both the charm and anticharm quarks were required to fall within the  $|\mathrm y| < 1.44$ rapidity window.
The distribution of the azimuthal angles between each pair ($\Delta\varphi$) was calculated to explore the event structures. 
In some of our results, we separately analysed the soft and hard production of the $\mathrm{c}$--$\mathrm{\overline{c}}$ pairs by requiring their momenta to fall below or above $p_{\rm T} = 4$ GeV/$c$.

We categorized the simulated events by the parton-level production process in which the trigger particles (charm quarks) were created. The flavour creation, flavour excitation and gluon splitting processes were separated in the simulations by tracing back the trigger charm quark to the first charm quark in the ancestry line and examining the status code of its parents. 
If this quark only had gluon parents that are not connected to the hardest process, the pair was considered to be from gluon splitting.
In the case when the charm quark had gluon parents, and it was an incoming particle in the hardest (sub)process, it was categorised as flavour excitation. 
When both of the parents were incoming light quarks or gluons in the hardest process, and created a charm quark, then the pair was categorised to come from pair creation. In a small number of events, where the charm did not originate from the hardest process, this method was not able to categorize the production process. These charm quarks, coming from subsequent soft processes, were added to the gluon splitting group.

The $\mathrm{c}$--$\mathrm{\overline{c}}$ azimuthal correlations were categorized with respect to charged hadron multiplicity, transverse spherocity and flattenicity. In all three cases, event variable cuts were applied to separate the top and bottom thirds of the sample.
Charged hadron multiplicity ($N_{\mathrm ch}$) was defined as the number of final charged hadrons with a transverse momentum of $p_{\rm T} > 0.15 $~GeV in the central pseudorapidity range $|\eta| < 1$. 
The low-activity range was taken as $N_{\rm ch} \leq 21 $, and the high-activity range as $N_{\rm ch} \geq 38$.
Transverse spherocity is calculated by finding the unit vector $\vec{n}$ that minimalises the expression
$$S_0 = \frac{\pi^2}{4} \left( \frac{\Sigma_i |\vec{p_{T_{i}}}\times \vec{n}|}{\Sigma_i |\vec{p_{T_{i}}}|} \right) ^2 ,$$
where the sum runs over all final-state charged particles with $p_{\rm T}>0.15$~GeV and $|\eta| < 1$.
With this definition, transverse spherocity is $0<S_0<1$, where $S_0 \approx 0$ events have a "pencil-like"  back-to-back dijet topology, and $S_0 \approx 1$ events are isotropic~\cite{Ortiz:2015ttf}.
For the low-$S_0$ range $S_0 < 0.53$ was used, and for the high-$S_0$ range $S_0 > 0.70$ was used.
Flattenicity ($\rho$) is a recently introduced event-shape variable that describes the distribution of transverse momenta over the azimuthal angle -- pseudorapidity plane,
which is capable of selecting "hedgehog-like" events without any discernible jetty structure in high-multiplicity pp collisions~\cite{Ortiz:2022zqr}. To calculate it, one has to divide the $\varphi$--$\eta$ plane into equal sections, and take the average transverse momenta of the charged particles in each of them. Flattenicity is the relative standard deviation of the average momentum in a cell:
$$\rho = \frac{\sigma _{p_{\rm T}^{\rm cell}}}{\langle p_{\rm T}^{\rm cell} \rangle} .$$
Larger $\rho$ implies a more jetty event, and around $\rho \approx 1$ is where at least one jet can be seen~\cite{Ortiz:2022zqr}.
The low flattenicity range was $\rho < 1.00 $, and the high flattenicity range was $\rho > 1.28$.

\section{Results}

First, we compare $\mathrm{c}$--$\mathrm{\overline{c}}$ azimuthal correlations for the high and low values of the event-shape variables, as well as without selection for the variables. The distributions are normalised with the number of triggers $N_{\rm trig}$, as well as with  the integral of the distributions for the given range divided by the integral of the distribution without selection for the event-shape variable, $I_{\rm class}$.

\begin{figure}
    \centering
    \begin{minipage}{0.4\linewidth}
    \includegraphics[width=\linewidth]{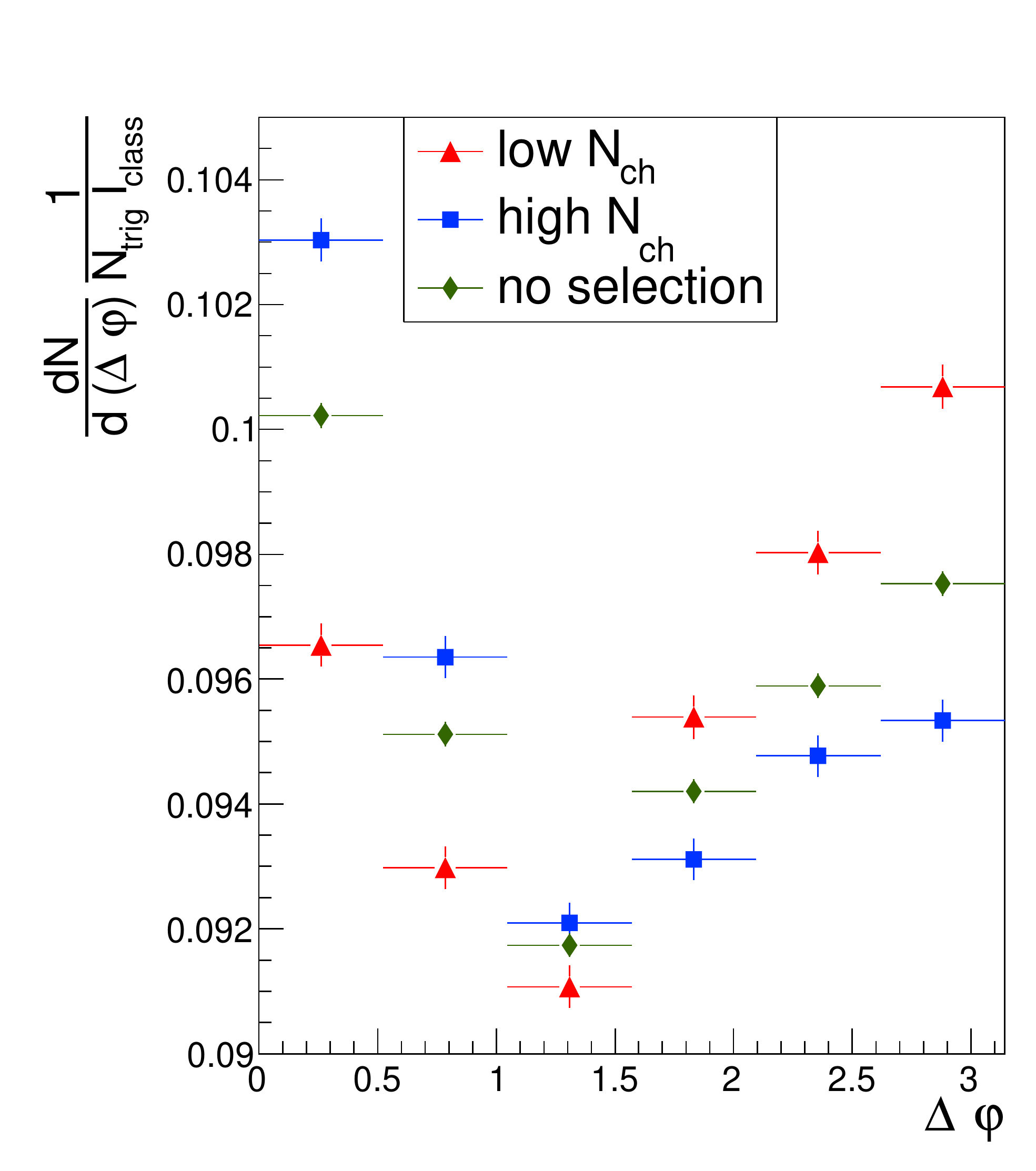}
    \caption{The azimuthal correlation of  $\rm c$--$\overline{\rm c}$ pairs in the low and high charged hadron multiplicity ranges, as well as without selection for $N_{\mathrm ch}$, normalised by the number of triggers and the integral of the interval.}
    \label{fig:nch}
    \end{minipage}
    \hspace{0.2cm}
    \begin{minipage}{0.4\linewidth}
    \centering
    \includegraphics[width=\linewidth]{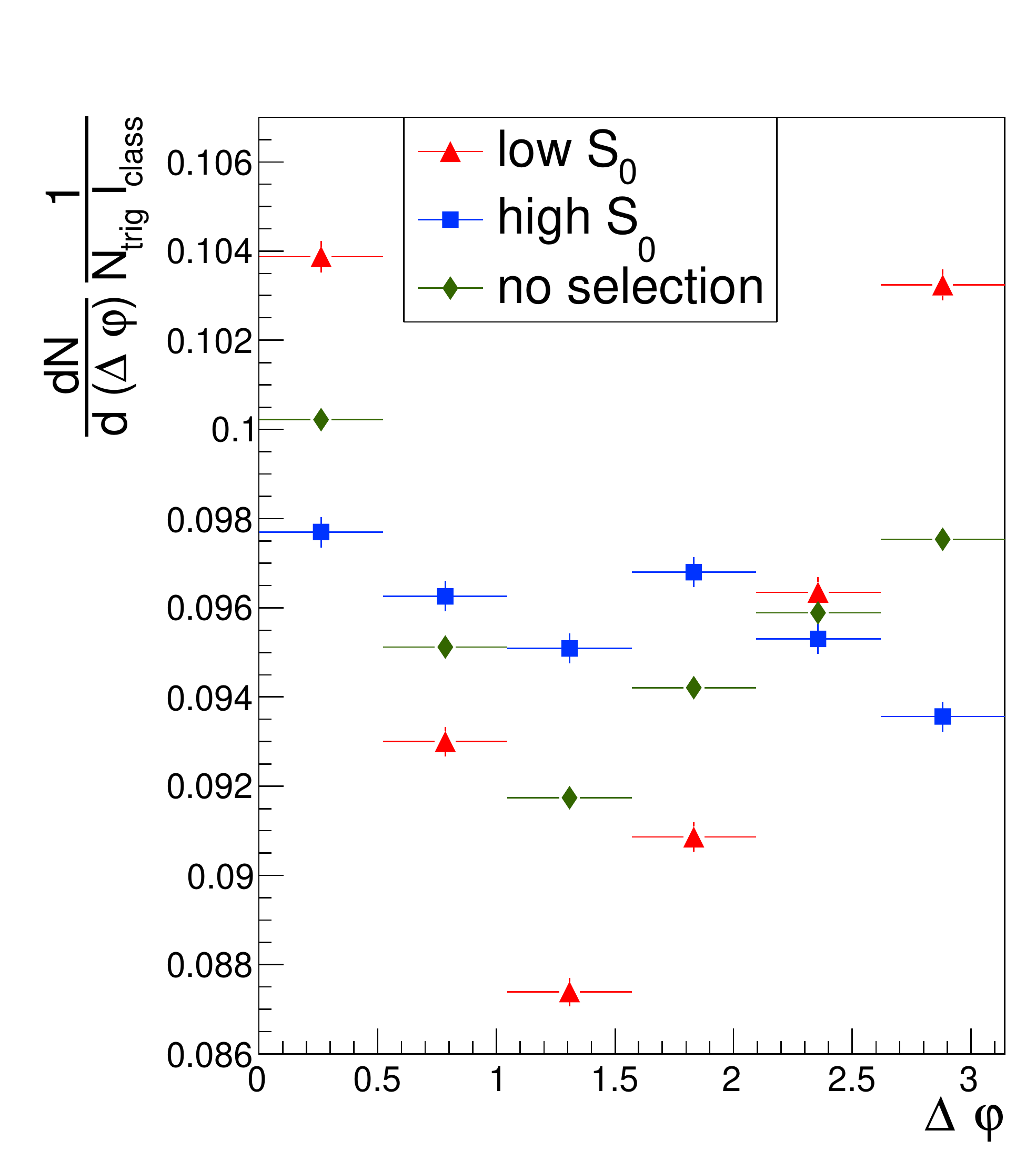}
    \caption{The azimuthal correlation of $\rm c$--$\overline{\rm c}$ pairs in the low and high transverse spherocity ranges, as well as without selection for $S_0$, normalised by the number of triggers and the integral of the interval.}
    \label{fig:s0}
    \end{minipage}
\end{figure}

Figure~\ref{fig:nch} shows the azimuthal correlation of $\rm c$--$\overline{\rm c}$ pairs in the low and high charged hadron multiplicity ranges as well as without selection for $N_{\mathrm ch}$.
We observe that for lower multiplicities the away-side peak of the correlation is sharper than at higher multiplicities. This can be explained by considering that low-multiplicity events are produced more often from simpler back-to-back correlations, while events with more complicated underlying physics tend to have higher multiplicities. 

Figure~\ref{fig:s0} shows the azimuthal correlation of $\rm c$--$\overline{\rm c}$  pairs in the low and high transverse spherocity ranges as well as without selection for $S_0$. We observe that events with low spherocity, which tend to be more jetty, result in a stronger correlation, and the more isotropic high $S_0$ range selects more random correlation.

\begin{figure}[h]
    \centering
    \begin{minipage}{0.4\linewidth}
    \centering
    \includegraphics[width=\linewidth]{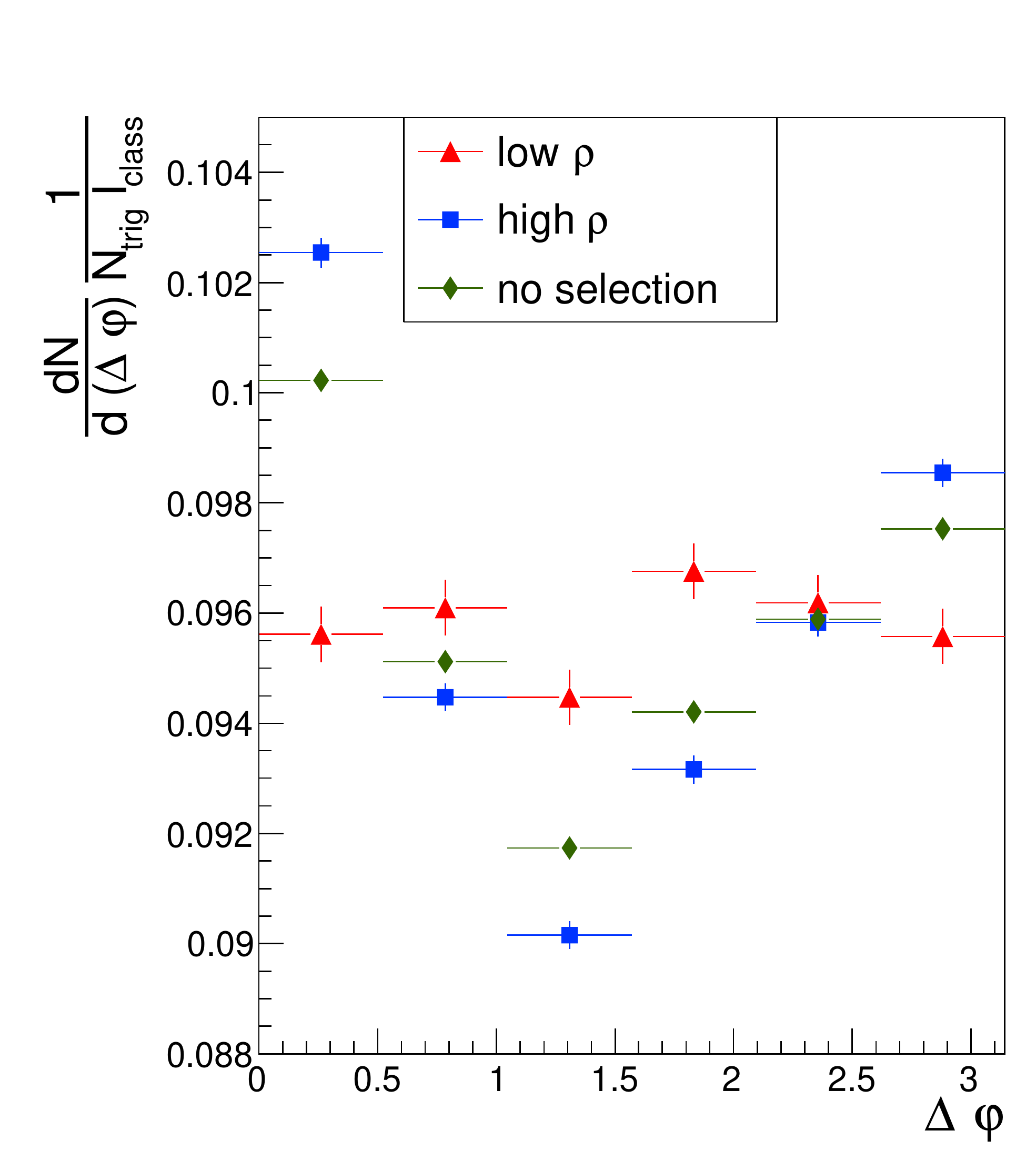}
    \caption{The azimuthal correlation of $\rm c$--$\overline{\rm c}$ in the low and high flattenicity ranges, as well as without selection for $\rho$, normalised by the number of triggers and the integral of the interval.}
    \label{fig:flat1}
    \end{minipage}
    \hspace{0.2 cm}
    \begin{minipage}{0.4\linewidth}
    \vspace{1.4 cm}
    \centering
    \includegraphics[width=\linewidth]{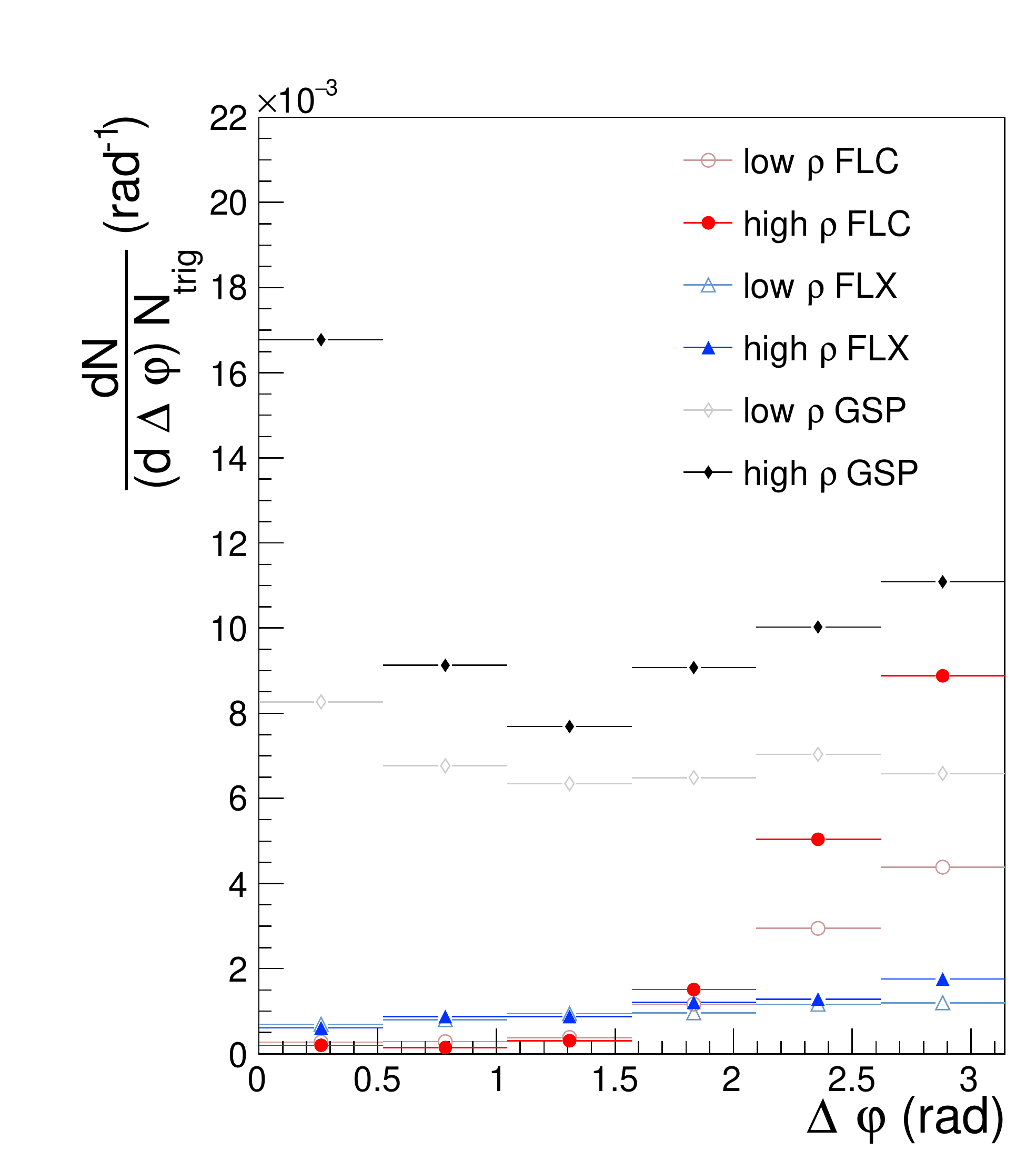}
    \caption{The azimuthal correlation of c--$\overline{\rm c}$ pairs separated by the different parton-level quark creation processes: flavour creation (in red), flavour excitation (in blue) and gluon splitting (in grey), in the $\rm c$ and $\overline{\rm c}$ momentum range of $p_{\rm T}>4$ GeV/$c$, in the low and high $\rho$ ranges (empty and full markers respectively).}
    \label{fig:flat3}
    \end{minipage}
\end{figure}

Figure~\ref{fig:flat1} shows the azimuthal correlation of $\rm c$--$\overline{\rm c}$  pairs in the low and high flattenicity cuts. We can see that flattenicity highlights the correlation peaks, as high $\rho$ gives both a sharper near-side and away-side peak, similarly to the observations made from the $S_0$ correlations.

In Figure~\ref{fig:flat3} we see the different trigger quark creation processes in the same flattenicity intervals, normalised by the number of triggers. Both the trigger and associated particle were required to have a transverse momentum $p_{\rm T}>4$ GeV/$c$. 
The dominant creation process in this high-$p_{\rm T}$ range is gluon splitting, which gives the largest contribution to the near-side peak, while adding to the away-side peak as well. Flavour creation gave a sharp away-side peak, and, though less visible, flavour excitation also adds mainly to the away-side peak. We can see that the flattenicity cut separates the peaks of gluon splitting (high $\rho$ GSP) from mostly random correlation (low $\rho$ GSP), which can be attributed to flattenicity geometrically separating isotropic events from jetty events.
We can also note that above a flat baseline of mostly gluon splitting, the away-side peak in the low $\rho$ range arises mainly due to flavour creation. On the other hand, the near-side peak in the high $\rho$ range is created by gluon splitting. We can see that flattenicity has the ability to geometrically separate these different creation processes via azimuthal correlation of c--$\overline{\rm c}$ quarks. This could provide an opportunity to experimentally separate different QCD production processes by observing the distribution of final-state particles through correlations of heavy-flavour jets.

\begin{figure}[ht]
    \centering
    \includegraphics[width=0.9\linewidth]{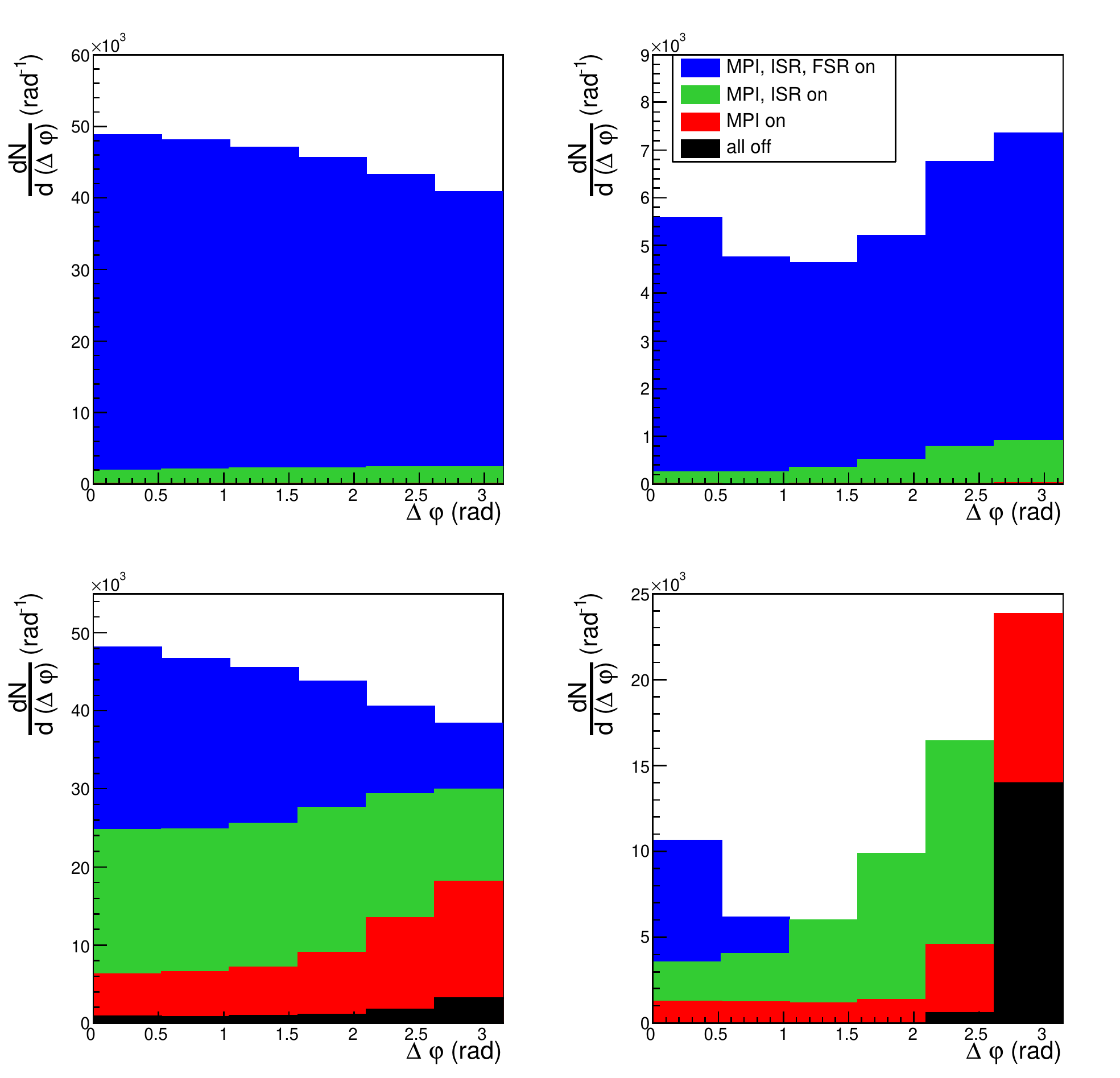}
    \caption{ The azimuthal correlation of c--$\overline{\rm c}$ pairs where  $p_{\rm T} < 4$ GeV/$c$ and $p_{\rm T} > 4$ GeV/$c$ (left and right columns respectively), and the top row shows the low $\rho$ range, and the bottom row shows the high $\rho$ range. The different parton-level process settings are presented with different colours.}
    \label{fig:flat4}
\end{figure}

Figure~\ref{fig:flat4}
shows the contributions of different PYTHIA8 parton-level processes for high and low $\rho$ values (top and bottom rows respectively), both in the $p_{\rm T} < 4$ GeV/$c$ and $p_{\rm T} > 4$ GeV/$c$ momentum ranges separately (left and right panels).
The c--$\overline{\rm c}$ pairs created back-to-back in the initial leading-order production result in an away-side peak. Multi-parton interactions and initial-state radiations also add to the away-side peak, while contributing to the baseline as well. The near-side peak arises from final-state radiations. 
Contrasting the two rows, we see that the flattenicity cut isolates most of the final-state radiation from multi-parton interaction and initial-state radiation. We also observe that the correlation peaks are stronger in the high $\rho$ events, that on average correspond to more jetty topologies. As expected, higher transverse momenta also results in less baseline.


\section{Conclusions}
In this work we explored the azimuthal correlation of charm quarks and antiquarks in PYTHIA8-simulated proton--proton collisions with respect to final-state charged hadron multiplicity, transverse spherocity and flattenicity. We investigated the event-activity and event-shape dependent results in terms of different QCD heavy flavour creation processes as well as parton-level processes. 
We observed that flattenicity is the most selective for the different QCD processes. 
By selecting events with low and high flattenicity, on a statistical basis we were able to differentiate between parton-level production processes, just by observing the event shape.
Moreover, by selecting low-flattenicity events we can also differentiate c--$\overline{\rm c}$ pairs coming from events with final-state radiation.

Using the above-mentioned methods it will be possible to select certain QCD processes in future heavy-quark (such as $\rm D^0$--$\overline{\rm D^0}$ or jet--jet) azimuthal correlation measurements in the LHC Run3 data. The results also outline a method for the detailed validation of heavy-flavour production models with data. 

\vspace{6pt} 


\authorcontributions{
Conceptualization, R.V.; methodology, R.V. and E.F.; software, A.H. and E.F.; validation, R.V.; formal analysis, A.H.; writing--original draft preparation, A.H.; writing--review and editing, R.V. and A.H.; visualization, A.H.; supervision, R.V.; project administration, R.V.; funding acquisition, R.V. All authors have read and agreed to the published version of the manuscript.
}

\funding{This work has been supported by the NKFIH grants OTKA FK131979 and K135515, as well as by the 2021-4.1.2-NEMZ\_KI-2022-00007 project.}

\dataavailability{
Simulated data is available from the Authors upon request.
} 

\acknowledgments{In this section you can acknowledge any support given which is not covered by the author contribution or funding sections. This may include administrative and technical support.}

\conflictsofinterest{
The authors declare no conflict of interest.
} 



\begin{adjustwidth}{-\extralength}{0cm}

\reftitle{References}


\bibliography{zimanyispec}

\begin{thebibliography}{999}

\bibitem[Adcox \em{et~al.}(2005)Adcox et~al.]{PHENIX:2004vcz}
Adcox, K.;  et~al.
\newblock {Formation of dense partonic matter in relativistic nucleus-nucleus
  collisions at RHIC: Experimental evaluation by the PHENIX collaboration}.
\newblock {\em Nucl. Phys. A} {\bf 2005}, {\em 757},~184--283,
  \href{http://xxx.lanl.gov/abs/nucl-ex/0410003}{{\normalfont
  [nucl-ex/0410003]}}.
\newblock {\url{https://doi.org/10.1016/j.nuclphysa.2005.03.086}}.

\bibitem[Adams \em{et~al.}(2005)Adams et~al.]{STAR:2005gfr}
Adams, J.;  et~al.
\newblock {Experimental and theoretical challenges in the search for the quark
  gluon plasma: The STAR Collaboration's critical assessment of the evidence
  from RHIC collisions}.
\newblock {\em Nucl. Phys. A} {\bf 2005}, {\em 757},~102--183,
  \href{http://xxx.lanl.gov/abs/nucl-ex/0501009}{{\normalfont
  [nucl-ex/0501009]}}.
\newblock {\url{https://doi.org/10.1016/j.nuclphysa.2005.03.085}}.

\bibitem[Khachatryan \em{et~al.}(2010)Khachatryan et~al.]{CMS:2010ifv}
Khachatryan, V.;  et~al.
\newblock {Observation of Long-Range Near-Side Angular Correlations in
  Proton-Proton Collisions at the LHC}.
\newblock {\em JHEP} {\bf 2010}, {\em 09},~091,
  \href{http://xxx.lanl.gov/abs/1009.4122}{{\normalfont
  [arXiv:hep-ex/1009.4122]}}.
\newblock {\url{https://doi.org/10.1007/JHEP09(2010)091}}.

\bibitem[Acharya \em{et~al.}(2019)Acharya et~al.]{ALICE:2019zfl}
Acharya, S.;  et~al.
\newblock {Investigations of Anisotropic Flow Using Multiparticle Azimuthal
  Correlations in pp, p-Pb, Xe-Xe, and Pb-Pb Collisions at the LHC}.
\newblock {\em Phys. Rev. Lett.} {\bf 2019}, {\em 123},~142301,
  \href{http://xxx.lanl.gov/abs/1903.01790}{{\normalfont
  [arXiv:nucl-ex/1903.01790]}}.
\newblock {\url{https://doi.org/10.1103/PhysRevLett.123.142301}}.

\bibitem[Bierlich \em{et~al.}(2018)Bierlich, Gustafson, and
  L\"onnblad]{Bierlich:2017vhg}
Bierlich, C.; Gustafson, G.; L\"onnblad, L.
\newblock {Collectivity without plasma in hadronic collisions}.
\newblock {\em Phys. Lett. B} {\bf 2018}, {\em 779},~58--63,
  \href{http://xxx.lanl.gov/abs/1710.09725}{{\normalfont
  [arXiv:hep-ph/1710.09725]}}.
\newblock {\url{https://doi.org/10.1016/j.physletb.2018.01.069}}.

\bibitem[Ortiz \em{et~al.}(2017)Ortiz, Bencedi, and Bello]{Ortiz:2016kpz}
Ortiz, A.; Bencedi, G.; Bello, H.
\newblock {Revealing the source of the radial flow patterns in
  proton\textendash{}proton collisions using hard probes}.
\newblock {\em J. Phys. G} {\bf 2017}, {\em 44},~065001,
  \href{http://xxx.lanl.gov/abs/1608.04784}{{\normalfont
  [arXiv:hep-ph/1608.04784]}}.
\newblock {\url{https://doi.org/10.1088/1361-6471/aa6594}}.

\bibitem[Norrbin and Sjostrand(2000)]{Norrbin:2000zc}
Norrbin, E.; Sjostrand, T.
\newblock {Production and hadronization of heavy quarks}.
\newblock {\em Eur. Phys. J. C} {\bf 2000}, {\em 17},~137--161,
  \href{http://xxx.lanl.gov/abs/hep-ph/0005110}{{\normalfont
  [hep-ph/0005110]}}.
\newblock {\url{https://doi.org/10.1007/s100520000460}}.

\bibitem[Ilten \em{et~al.}(2017)Ilten, Rodd, Thaler, and
  Williams]{Ilten:2017rbd}
Ilten, P.; Rodd, N.L.; Thaler, J.; Williams, M.
\newblock {Disentangling Heavy Flavor at Colliders}.
\newblock {\em Phys. Rev. D} {\bf 2017}, {\em 96},~054019,
  \href{http://xxx.lanl.gov/abs/1702.02947}{{\normalfont
  [arXiv:hep-ph/1702.02947]}}.
\newblock {\url{https://doi.org/10.1103/PhysRevD.96.054019}}.

\bibitem[V\'ertesi \em{et~al.}(2021)V\'ertesi, Benc\'edi, Mis\'ak, and
  Ortiz]{Vertesi:2021ihp}
V\'ertesi, R.; Benc\'edi, G.; Mis\'ak, A.; Ortiz, A.
\newblock {Probing the interaction of semi-hard quarks and gluons with the
  underlying event in light- and heavy-flavor triggered proton-proton
  collisions}.
\newblock {\em Eur. Phys. J. A} {\bf 2021}, {\em 57},~301,
  \href{http://xxx.lanl.gov/abs/2107.06764}{{\normalfont
  [arXiv:hep-ph/2107.06764]}}.
\newblock {\url{https://doi.org/10.1140/epja/s10050-021-00602-9}}.

\bibitem[Ortiz and Paic(2022)]{Ortiz:2022zqr}
Ortiz, A.; Paic, G.
\newblock {A look into the \textquotedblleft{}hedgehog\textquotedblright{}
  events in pp collisions}.
\newblock {\em Rev. Mex. Fis. Suppl.} {\bf 2022}, {\em 3},~040911,
  \href{http://xxx.lanl.gov/abs/2204.13733}{{\normalfont
  [arXiv:hep-ph/2204.13733]}}.
\newblock {\url{https://doi.org/10.31349/SuplRevMexFis.3.040911}}.

\bibitem[Ortiz \em{et~al.}(2015)Ortiz, Pai\'c, and Cuautle]{Ortiz:2015ttf}
Ortiz, A.; Pai\'c, G.; Cuautle, E.
\newblock {Mid-rapidity charged hadron transverse spherocity in pp collisions
  simulated with Pythia}.
\newblock {\em Nucl. Phys. A} {\bf 2015}, {\em 941},~78--86,
  \href{http://xxx.lanl.gov/abs/1503.03129}{{\normalfont
  [arXiv:hep-ph/1503.03129]}}.
\newblock {\url{https://doi.org/10.1016/j.nuclphysa.2015.05.010}}.

\bibitem[Acharya \em{et~al.}(2022)Acharya et~al.]{ALICE:2021kpy}
Acharya, S.;  et~al.
\newblock {Investigating charm production and fragmentation via azimuthal
  correlations of prompt D mesons with charged particles in pp collisions at
  $\mathbf {\sqrt{ s} = 13}$ TeV}.
\newblock {\em Eur. Phys. J. C} {\bf 2022}, {\em 82},~335,
  \href{http://xxx.lanl.gov/abs/2110.10043}{{\normalfont
  [arXiv:nucl-ex/2110.10043]}}.
\newblock {\url{https://doi.org/10.1140/epjc/s10052-022-10267-3}}.

\bibitem[Sjostrand \em{et~al.}(2008)Sjostrand, Mrenna, and
  Skands]{Sjostrand:2007gs}
Sjostrand, T.; Mrenna, S.; Skands, P.Z.
\newblock {A Brief Introduction to PYTHIA 8.1}.
\newblock {\em Comput. Phys. Commun.} {\bf 2008}, {\em 178},~852--867,
  \href{http://xxx.lanl.gov/abs/0710.3820}{{\normalfont
  [arXiv:hep-ph/0710.3820]}}.
\newblock {\url{https://doi.org/10.1016/j.cpc.2008.01.036}}.

\end{thebibliography}

\PublishersNote{}
\end{adjustwidth}
\end{document}